\documentclass[final,1p,times,authoryear]{elsarticle}
\usepackage[T1]{fontenc}
\usepackage{ae,aecompl}
\usepackage{float}
\usepackage{booktabs}
\usepackage{amsmath}
\usepackage{bm}
\usepackage{soul}
\usepackage{graphicx}   % Including figure files
\newcommand{\jcap}{JCAP}

\newcommand{\apjs}{Astrophys. J. Suppl. Ser.}
\newcommand{\aap}{Astron. \& Astrophys.}
\newcommand{\aj}{Astron. J.}
\newcommand{\apj}{Astron. J.}
\newcommand{\nat}{Nature}
\newcommand{\sovast}{Soviet Astronomy}
\newcommand{\araa}{Ann. Rev. Astron. Astrophys. } %ARA$\&$A}
\newcommand{\mnras}{Mon. Not. R. Astron. Soc.}
\newcommand{\apss} {Astrophys. and Space Science}

\newcommand{\physrep}{Physics Reports}
\usepackage{ulem}
\newcommand{\rthis}[1]{\textcolor{black}{#1}}
\usepackage{amsmath}    % Advanced maths commands
\usepackage{amssymb}    % Extra maths symbols
\usepackage{color}      
\newcommand{\simgt}{\hbox{\,\rlap{\raise 0.425ex\hbox{$>$}}\lower 0.65ex\hbox{$\sim$}\,}}
\newcommand{\simlt}{\hbox{\,\rlap{\raise 0.425ex\hbox{$<$}}\lower 0.65ex\hbox{$\sim$}\,}}

\usepackage{natbib}
\usepackage[plainpages=false, colorlinks=true, anchorcolor=blue, linkcolor=blue, citecolor=blue, bookmarks=false]{hyperref}

\journal{New Astronomy}
\begin{document}
\begin{frontmatter}
\title{Classification of Pulsars using Extreme Deconvolution}
\author[a]{Tarun Tej Reddy Ch.}
%\altaffiliation{E-mail: ee17btech11042@iith.ac.in}

\author[b]{Shantanu Desai\corref{cor1}}
\cortext[cor1]{Corresponding Author}
\ead{shntn05@gmail.com}
%\altaffiliation{E-mail:shntn05@gmail.com}
 
 \address[a]{Department of Electrical Engineering, IIT Hyderabad, Kandi, Telangana 502285, India}
\address[b]{Department of Physics, Indian Institute of Technology, Hyderabad, Kandi, Telangana-502285, India} 

\begin{abstract}
We carry out a classification of  the observed pulsar dataset into distinct clusters, based on  the $P-\dot{P}$ diagram, using Extreme Deconvolution based Gaussian Mixture Model. We then use the Bayesian Information Criterion to select the optimum number of clusters. We find in accord with previous works, that the pulsar dataset can be optimally classified into  six clusters, with two for the millisecond pulsar population, and four for the ordinary pulsar population. Beyond that, however we do not glean any additional insight into the pulsar population based on this classification. Using numerical experiments, we confirm that Extreme Deconvolution-based classification is less sensitive to variations in the dataset compared to ordinary Gaussian Mixture Models. All our analysis codes used for this work have been made publicly available.
\end{abstract}

\end{frontmatter}
\section{Introduction}  
Neutron stars most commonly manifest themselves as radio pulsars.
Pulsars are rotating neutron stars which emit pulsed radio emission. The zeroth-order model to explain the pulsed radio emission  in pulsars is attributed to a rotating magnetic dipole, although the full  details of the pulsar emission are much more complex and still not completely understood~\citep{Melrose20}. Ever since their first serendipitous discovery of a pulsar (neutron star)~\citep{Hewish}, a whole zoo of  neutron stars with considerable diversity have been discovered throughout the electromagnetic spectrum.  Pulsars have proved  to be wonderful laboratories for a wide range of topics in   Physics  and Astronomy~\citep{handbook}. A few representative examples of these myriad connections of pulsars to the rest of Physics/Astrophysics include:
Solid state physics~\citep{bhattacharya}, Plasma Physics and  Fluid Mechanics~\citep{Blandford}, QED~\citep{Meszaros}, QCD~\citep{Alford} tests of General Relativity~\citep{Stairs03}, study of nuclear matter at high densities~\citep{Lattimer,Bagchi}, indirect probes of gravitational waves~\citep{Taylor}, exoplanets~\citep{Frail}, dark matter~\citep{Desai}, stellar evolution~\citep{Stairs04}, probes of interstellar medium~\citep{Frail94,Keith}, solar wind~\citep{Krishnakumar} etc.

The two main observables which typically characterize a radio pulsar are its period ($P$) and period derivative ($\dot{P}$). From these, one can obtain approximate estimates of their ages and magnetic fields. Radio pulsars can be broadly classified into two types. The first category is  ordinary radio pulsars with  periods greater than approximately 100 ms, $\dot{P}$ between $10^{-16}$ and $10^{-13}$, and magnetic fields between   $10^{10}-10^{13}$ G. Most of these pulsars are in isolated systems and the youngest of them are  usually associated with supernova remnants.
The second category is millisecond pulsars, with periods shorter  than approximately 100 ms and characteristic surface magnetic field $<10^{10}$ G. The origin of these pulsars is different from the classical radio pulsar population, and these are generally accepted to be the descendants  of low-mass X-ray binaries~\citep{bhattacharya}. More than  250  millisecond pulsars have also been discovered in gamma rays by the Fermi-LAT satellite, some of which are new discoveries, and have not yet been detected in the radio~\citep{Acero}. As of now, the total number of known pulsars is close to 3,000 and we expect to discover about 20,000 new pulsars in the SKA era~\citep{Strappers}.

Subsequently,  a whole zoo of neutron stars  with diverse characteristics~\citep{Konar,Kaspi10,Harding} have since been discovered, such as magnetars~\citep{Woods}, rotating radio transients~\citep{Mclaughlin06}, X-ray dim isolated neutron stars~\citep{Sandro}, Fast Radio bursts~\citep{Cordes}, Central compact objects~\citep{DeLuca}. Because of the proliferation of distinct observational classes of neutron stars,  a number of automated techniques have been used to classify the pulsar population into distinct classes, in the hope of gaining new insights into the connection between these classes. The first class of techniques involve the study of  evolutionary tracks along the $P$-$\dot{P}$ diagram, also known as  ``pulsar current'' analysis~\citep{Vivek,Melrose,Glushak}.
The second group of efforts involves the application of unsupervised clustering techniques. The first work along these lines was carried out by~\citet{kjlee}, who applied the Gaussian mixture model (GMM, hereafter)  in the $P-\dot{P}$ plane to  radio pulsars from the ATNF catalog~\citep{ATNF}.  
They found a total of six clusters (two for the millisecond pulsar population and four for the ordinary pulsars).
However,~\citeauthor{Popov} pointed out  that for neutron stars with evolution in  the $P-\dot{P}$ plane, GMM is not effective in distinguishing between such groups. They also found that the positions of the clusters gets changed when a random 10\% of the data is excluded.  Hence, they concluded that GMM is oversensitive to the pulsar dataset and does not produce stable results.  
~\citet{dirichlet} then used Dirichlet process GMM to classify a stacked catalog of neutron stars, which included   radio pulsars from the ATNF catalog, along with     SGRs, AXPs, RRATs, CCOs, and XDINS.  Their analysis also confirms the presence of two clusters for the millisecond pulsar population and four for the rest. 

In this work, for the purpose of pulsar classification, we incorporate the uncertainties in the observed  $P$ and $\dot{P}$,  and use  an extension of GMM, known as Extreme Deconvolution~\citep{Bovy}. The outline of this paper is as follows. A brief discussion of Extreme Deconvolution based GMM is given in Sect.~\ref{sec:xdgmm}. Details of our analysis and results can be found in Sect.~\ref{sec:analysis}. We conclude in Sect.~\ref{sec:conclusions}. All our codes used for this analysis have been made publicly available, for which the relevant link can be found in Sect.~\ref{sec:conclusions}. %Throughout the paper, log refers to $\log_{10}$.

\section{Extreme Deconvolution}
\label{sec:xdgmm}
The problem of density estimation and  finding clusters in a given dataset has widespread applications throughout Astrophysics. For this purpose,  a large number of techniques involving unsupervised clustering have been applied to a variety of problems~\citep{Brunner,astroML,Chatto,Fluke}. Here, ``unsupervised'' refers to the case, where there are no class labels. Unsupervised clustering techniques use all the available data to find the optimum number of classes.

A large class of these clustering algorithms involving Parametric density estimation come under the  guise of ``Mixture models''~\citep{astroML,Kuhn}, where a mixture model combines multiple components of probability distributions into a single one. The most widely used mixture models use Gaussian components,  and hence is called   Gaussian Mixture Model (GMM). 
%Although mixture models with non-Gaussian distributions have also been used in astrophysical literature~\citep{Tarnopolski}, the central limit theorem assures us that the  assumption of Gaussianity within each cluster is a reasonable one for any mixture model. 
As pointed out in ~\citeauthor{Popov}, since the logarithms of the magnetic field and age   are within a narrow range, this in turn implies a narrow range for their  variance.  The central limit theorem states that the mean of a large number of independent random variables with similar means and variances asymptotes to a normal distribution. Therefore, it is reasonable to assume Gaussianity to analyze the $P-\dot{P}$, in log-space as they are proportional to   the ages and magnetic fields (cf. Eq.~\ref{eq8} and Eq.~\ref{eq9}.) We however note that Gaussianity would be a poor approximation if we use the raw $P$ and $\dot{P}$ values. 

In GMM, the data is modelled by fitting it with a weighted sum of  multiple Gaussian distributions,  with each component having separate means and covariance.   GMM has been widely used for a plethora of classification problems in astrophysics~\citep{Kuhn}. 
Inferring distributions of the data which involve uncertainties is even more challenging compared to data without uncertainties, as the noise for each observation could be generated by an unknown source. This extension of GMM which incorporates the uncertainties in the data is known in the Astrophysics literature as Extreme Deconvolution (XDGMM)~\citep{Bovy,astroML,Wechsler}.  We first provide a brief mathematical preview of GMM, and then outline how it can be generalized  to XDGMM by including the uncertainties.  More details on XDGMM are provided in the  aforementioned works.

The  GMM models the distribution  as a mixture of Gaussian clusters. Each Gaussian cluster has a weight, a central data point (mean) and a covariance matrix  associated with it. The likelihood of each data point ($\mathbf{x}$) for a GMM is given by: 
\begin{equation}
\begin{split}
p(\mathbf{x}) & = \sum_{j=1}^{K} \alpha_j \mathcal{N}(\mathbf{x}|\bm{\mu_j},\Sigma_j),  \text{where} \\
\sum_{j=1}^{K} \alpha_j &= 1
\end{split}
\label{eq:gmm}
\end{equation}
where  $\alpha_j$, $\bm{\mu_j}$,  and $\Sigma_j$ are the weights, means, and covariance matrix of  the $j^{th}$ Gaussian cluster,  $\mathcal{N}(\mathbf{x}|\bm{\mu_j},\Sigma_j)$ is the Gaussian probability density function of the $j^{th}$ Gaussian cluster, $K$ is the total number of clusters. We note that $\mathbf{x}$ and $\bm{\mu_j}$ are in general, vectors. For each data point, one can define a class probability ($p(j|\mathbf{x})$), that it was generated by the class $j$:
\begin{equation}
p(j|\mathbf{x}) = \frac{\alpha_j  \mathcal{N}(\mathbf{x}|\bm{\mu_j},\Sigma_j)}{p(\mathbf{x})}   
\end{equation}
where $p(\mathbf{x})$ is defined in Eq~\ref{eq:gmm}.
The best-fit parameters for each of the clusters are found using Expectation-Maximization (E-M) algorithm~\citep{EM}.  This algorithm exploits the fact that the class probability is known and fixed in each iteration. Therefore, the derivative of the log-likelihood reduces to a simple algebraic function of the means and variances of each of the Gaussian components. These can then be determined iteratively. More details about the E-M algorithm can be found in ~\citet{EM} and references therein.

XDGMM now generalizes the original GMM, by taking into account the uncertainty distribution of each data point.  We assume that the noisy dataset $x_i$ is related to the true values $v_i$ as follows~\citep{Bovy,astroML}:
\begin{equation}
x_i=R_i v_i + \epsilon_i,
\label{eq:transform}
\end{equation}
where $R_i$ is the rotation matrix used to transform the true values to the observed noisy dataset. In this particular case $R_i$ is the identity matrix, because we measure  $P$ and $\dot{P}$ directly.
Note that $x_i$
and $v_i$ could be multi-dimensional vectors and for our example, denote the 2-D dataset comprising of $P-\dot{P}$ for the pulsars.
The noise $\epsilon_i$ is assumed to be drawn from a Gaussian with zero mean and variance $S_i$.  Then, the likelihood of the model parameters ($\theta$={$\alpha$, $\mu$, $\Sigma$}) for each data point is given as,\\
\begin{equation}
p(x_i|\theta)=\sum_{j=1}^{K} \alpha_j \mathcal{N}(x_i|R_i\mu_j,R_i\Sigma_jR_i^{T}+S_i)
\end{equation}

The final step is to maximize the likelihood of the dataset with respect to the model parameters. This can be done (as in GMM) by summing up  the individual log-likelihood functions. 
\begin{equation}
\phi=\sum_{i=1}^{N} \ln (p(x_i|\theta)),
\end{equation}
where $N$ is the total number of datapoints.
Similar to GMM, a simple extension  of the   EM   algorithm (discussed in ~\citet{Bovy}) is used to maximize the objective function in XDGMM.  The EM algorithm iteratively maximizes the likelihood, and hence results in the optimal values of the model parameters. 

\iffalse
The expectation step is given by,
\begin{equation}
\begin{split}
 q_{ij} & \leftarrow \frac{\mathcal{N}(x_i|R_i\mu_j,R_i\Sigma_jR_i^{T}+S_i)}{\sum_{k} \alpha_k \mathcal{N}(x_i|R_i\mu_k,R_i\Sigma_kR_i^{T}+S_i)} \\
    b_{ij} & \leftarrow \mu_j+\Sigma_jR^T(R_i\Sigma_kR_i^{T}+S_i)^{-1}(x_i-R_i\mu_j)\\
    B_{ij} & \leftarrow \Sigma_j-\Sigma_jR_i^T(R_i\Sigma_kR_i^{T}+S_i)^{-1}R_iV_j
\end{split}
\end{equation}
\newline
The maximization step is given by,
\begin{equation}
\begin{split}
\alpha_j & \leftarrow \frac{1}{N} \sum_{i}q_{ij}\\
    \mu_j & \leftarrow \frac{1}{q_j}\sum_{i}q_{ij}b_{ij}\\
    \Sigma_j & \leftarrow \frac{1}{q_j}\sum_{i}q_{ij}[(\mu_j-b_{ij})(\mu_j-b_{ij})^T+B_{ij}]
\end{split}
\end{equation}

Here $q_j$=$\sum_{i}q_{ij}$ and $b_{ij}$ and $B_{ij}$ are the conditional distribution and covariance matrix respectively. Repeating the above expectation and maximization steps iteratively will give the optimal weights, means and the variances of the Gaussian clusters.
\fi

XDGMM has proven to be useful in modelling the underlying distributions, where the data points have uncertainties associated with them, such as the velocity distribution from Hipparcos data~\citep{Bovy},  the three-dimensional motions  of the stars in Sagittarius streams ~\citep{sagittarius_koposov}, classification of neutron star masses~\citep{Keitel}, identification of dark matter subhalo candidates~\citep{Miguel}.

\section{Analysis and Results}
\label{sec:analysis}
We now apply the XDGMM algorithm to classify the  pulsar  population in the $P$-$\dot{P}$ plane. First, we describe the dataset used  for the analysis, followed by the implementation  of the XDGMM algorithm. We then discuss the metric used for choosing the optimum  number of Gaussian components, and finally present our results.

\subsection{Data Collection} \label{data}
For this work, we download  $P$ and $\dot{P}$ for the radio pulsar population, along with the measurement uncertainties  from the ATNF online catalogue (version \rthis{1.65})~\cite{ATNF}~\footnote{ \url{http://www.atnf.csiro.au/research/pulsar/psrcat/}}. \rthis{We used the values from the field {$\tt P1\_i$}, whenever they were available, else {\tt P1} was used. The {$\tt P1\_i$} values are relevant for the millisecond pulsar population.} The ATNF catalog contains an up-to-date list of all the  discovered pulsars. At the time of writing, there were exactly 2,374 pulsars with known uncertainties in $P$ and $\dot{P}$ and positive values for $\dot{P}$.  This catalog also contains \rthis{756} pulsars with either negative values for $\dot{P}$ or missing uncertainties for $P$ or $\dot{P}$. We note that the measured  $\dot{P}$ can differ from the intrinsic $\dot{P}$, because some sources are being significantly accelerated in the gravitational potential of the galaxy or their host globular cluster. Another reason which is important for millisecond pulsars is the  Shklovskii effect~\citep{Shlov} due to the transverse motion of the pulsar across the sky.
 All these factors which affect the intrinsic pulsar periods have been recently reviewed in ~\cite{Pathak}. 
These 756 pulsars were excluded from our analysis. %The distribution of $P$ and $\dot{P}$ for these 2289 pulsars is shown in Fig.~\ref{fig1}. 
Unlike ~\cite{kjlee,dirichlet}, we did not include ancillary neutron stars  such as RRATs, CCO, gamma-ray pulsars (from Fermi-LAT), since the  error estimates for $P$ and $\dot{P}$ for these datasets were not  available. However, it is trivial to include these in our analysis, if their error estimates are made available.
%\begin{figure}[H]
%\centering
%    \includegraphics[scale=0.55]{P_Pdot_plot.png}
%   \caption{Pulsar P-$\dot{P}$ diagram using the online ATNF catalog~\citep{ATNF}.}
%   \label{fig1}
%\end{figure}

\iffalse
The errors in $P$ and $\dot{P}$ are represented as $P_e$ and $\dot{P}_e$ respectively. Fig.~\ref{fig2} shows the pulsar $P$-$\dot{P}$ diagram including the errors in $P$ and $\dot{P}$.
\begin{figure}
\centering
\includegraphics[scale=0.55]{P_Pdot_plot_error.png}
\caption{Pulsar $P$-$\dot{P}$ diagram with errors in $P$ and $\dot{P}$. The red bars represent the errors in $P$ and $\dot{P}$.}
\label{fig2}
\end{figure}
\fi

 \subsection{XDGMM for Pulsar Classification} 

We now  apply XDGMM to the pulsar dataset. Since both the $P$  and $\dot{P}$ values have a large dynamic range spanning many orders of magnitude, the inputs given to the XDGMM algorithm  are $\ln(P)$
and $\ln(\dot{P})$, similar to ~\cite{kjlee,dirichlet}. The errors in these transformed variables are obtained by error propagation.

\iffalse
\begin{equation}
\begin{split}
 P \rightarrow \ln(P)  \\
\dot{P} \rightarrow \ln(\dot{P}) \end{split}
\end{equation}

Since the $P$ and $\dot{P}$ are converted to a log scale, the errors in $P$ and $\dot{P}$ will change too. The propagation of error for different functions can be found using Eq. \ref{eq8err}.\\
\begin{equation} \label{eq8err}
\sigma(f(x))=f^{'}(x)*\sigma(x)
\end{equation}\\

Here $x$ is a data point taken from a one dimensional data set. $f(x)$ is a continuous function on x, $\sigma$(f(x)) is the error in f(x) and $\sigma$(x) is the error in x. In the case of pulsar data, $P$ and $\dot{P}$ values form a one dimensional data set each. Since $P$ and $\dot{P}$ are converted to their logarithmic values, f(x) is taken as log(x). Therefore, using Eq.~\ref{eq8}, the errors in $P$ and $\dot{P}$ are obtained as,\\
\begin{equation}
\begin{split}
 P_e \rightarrow P_e/log(P)  \\
\dot{P}_e \rightarrow \dot{P}_e/log(\dot{P}) \end{split}
\end{equation}
\newline
\fi

For this work, we used the implementation of XDGMM  from the  {\tt astroML} python module~\citep{astroML}. The dataset provided as input to the XDGMM code consists of $\ln(P)$ and $\ln(\dot{P})$.  Since,   we need to classify the pulsars using a 2-d dataset, we  vertically stack the $\ln(P)$ and $\ln(\dot{P})$ values, and provide the resulting matrix as input to the XDGMM algorithm. Therefore, we get two error covariance matrices for every 2-d datapoint taken from $\ln(P)$ and $\ln(\dot{P})$. The uncertainties in $\ln(P)$  and $\ln(\dot{P})$ constitute the diagonal elements of their respective covariance matrices, with non-diagonal elements kept at zero, since we assume that the errors between the different pulsars are uncorrelated. We note however that this assumption  may  not be correct all the time,   since  the errors are related to the capability of the used pulsar detection system and each of these systems is usually responsible for the $P$ and $\dot{P}$ values (and uncertainties on those) for many pulsars.  Thus, not all the estimated uncertainties are indeed uncorrelated, especially for pulsars discovered from the same survey. However a detailed characterization of these covariances is hard to model, given the large number of pulsars detected through multiple  hetereogeneous surveys, and is  beyond the scope of this work.
These covariance matrices are vertically stacked and the resulting matrix is provided as an input  to the XDGMM algorithm.  After running XDGMM, one can obtain the weights, means, and covariances for the specified number of clusters.

The number of Gaussian components used to fit this data is very important. We must ensure that the models do not  underfit or overfit the data.  Similar to most mixture models, XDGMM by itself does not determine the optimum number of clusters, and these are usually provided as inputs to the algorithm.  In ~\citet{kjlee} (and also in ~\citet{Popov}), the optimum number of clusters were determined using a 2-D Kolmogorov-Smirnov (K-S) test.  However, concerns about the  validity of the 2-D K-S test have been raised in literature~\citep{Babu}~\footnote{ \url{https://asaip.psu.edu/articles/beware-the-kolmogorov-smirnov-test/}}. Here, we treat the determination of optimum number of clusters as a model selection problem, and choose  an information theory based metric, similar to our past work on GMM-based classification of GRBs and exoplanets~\citep{Kulkarni,Kulkarniexo}.

\subsection{Bayesian  Information  Criterion } \label{BIC}
The Bayesian Information Criterion ~\citep{schwarz,Liddle} score is an approximation to Bayesian evidence. BIC compensates for additional free parameters, and is widely used in astrophysics and cosmology for model comparison.  The equation for BIC can be written as:
\begin{equation}
    BIC = -2\ln(\hat{L})+k \ln(n),
\end{equation}
\label{eq7}
where $\hat{L}$ is the maximum likelihood  of a given model, $n$ is the size of the data set, and $k$ is  the total number of free parameters to be used to fit the model. BIC penalizes for additional number of free parameters, and hence aids in rejecting models which overfit the data. While comparing two models, the one with the lower BIC score is chosen as the optimum one.

\begin{figure}[H]
\centering
\includegraphics[scale=0.55]{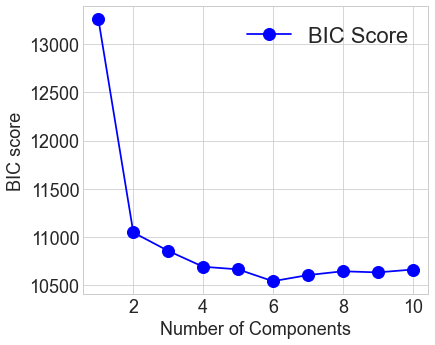}
\caption{BIC Scores for the different number of Gaussian clusters used to fit the $P$ and $\dot{P}$ data.}
\label{fig3}
\end{figure}

For our analysis, we apply XDGMM with different number of clusters as inputs. For each of these choices, we compute the BIC score. These BIC scores as a function of the number of clusters used for classifying the pulsar data are  plotted in Fig.~\ref{fig3}. We find a minimum value of BIC for six clusters. Therefore, the optimum number of Gaussian clusters needed to fit the pulsar  $P$-$\dot{P}$ data including their errors  is equal to six. This agrees with the analysis in ~\citet{kjlee} and \citet{dirichlet}, who also found six clusters using GMM and Dirichlet-GMM, respectively.

\subsection{Results}
The resulting weights, means, and covariances of the six different Gaussian clusters, which can be used to classify the pulsar population in the $P$-$\dot{P}$ logarithmic plane  are tabulated in Table~\ref{Table1}. The clusters A, B and E have the same weights, means and covariances as that  reported in ~\cite{kjlee}. Cluster F has the  same weight as in ~\cite{kjlee} but for clusters C and D, the weights have changed by 15\%. Cluster D has the same mean and covariance as reported in ~\cite{kjlee}. The means of clusters C and F vary by nearly 0.5 seconds (in log units) in the $x$ direction and 0.4 (in log units) in the $y$ direction.  The lengths of semi-major and semi-minor axis of the Clusters C and F differ by 0.5 and 0.3 (both in log scale), respectively.
Only one cluster has a mean with a positive value for $\ln(P)$, and only one covariance matrix has negative non-diagonal values. These observations are therefore in accord with  the corresponding means and covariances obtained in ~\cite{kjlee}, who had also analyzed the radio pulsar sample from the ATNF catalog (circa 2012).

%------------------------------------------

\begin{figure*}
\centering
\includegraphics[scale=0.8]{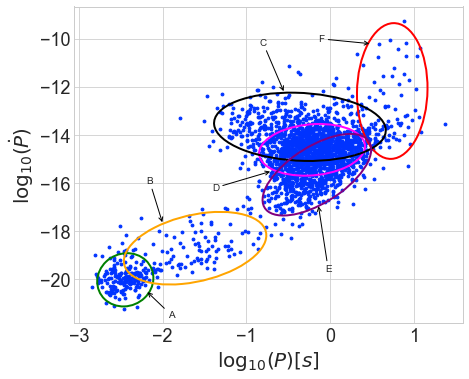}
\caption{Pulsar P-$\dot{P}$ diagram with Gaussian clusters, obtained after applying XDGMM. The ellipses denote the 95\% c.i.  contours for the six Gaussian clusters obtained from XDGMM. Clusters A and B describe the millisecond pulsars, whereas the remaining clusters encapsulate the rest of the radio pulsar population. More details about their properties can be found in Table~\ref{Table1}.}
\label{fig4}
\end{figure*}

The  means and covariances of the Gaussian clusters are used to plot 95\% confidence interval (c.i.) ellipses in the $P$-$\dot{P}$ plane. The corresponding plots obtained by superimposing these confidence ellipses in the $P$-$\dot{P}$ diagram are shown in Fig.~\ref{fig4}. The two Gaussian clusters corresponding to the millisecond pulsars (MSPs) are independent of the clusters with high values for the period and larger magnetic fields.  More details on the difference between these two sets of millisecond pulsars is discussed in \citet{kjlee, dirichlet}. We note however that there are exceptions to the binary types  corresponding to these two sets pointed in the aforementioned works.
The remaining four Gaussian clusters correspond to the ordinary pulsar population. We however note that there is also a diversity in the pulsar population in each of the  clusters identified by XDGMM. 

From the pulsar $P$ and $\dot{P}$, one can estimate a characteristic age ($\tau_c$) and the characteristic surface magnetic field ($B_{s}$) after making certain assumptions~\citep{handbook,dirichlet}:
\begin{equation}
\tau_c = \frac{P}{2\dot{P}}  
\label{eq8}
\end{equation}
\begin{equation}
B_s=\left(\frac{3c^3I}{8\pi^2R^6}\right)^{\frac{1}{2}}\sqrt{P\dot{P}}
\label{eq9}
\end{equation}
where $I$ is the moment of inertia and $R$ is the radius of the neutron stars. Similar to  ~\cite{dirichlet}, we  assumed $I=10^{45} \rm{g cm^2}$ and $R=10^6$ cm. Therefore Eq.~\ref{eq9} simplifies to, $B_s=3.2\times 10^{19} \sqrt{P\dot{P}}$ G.  

From these equations, we then proceed to calculate the characteristic age and surface magnetic field for each of the cluster populations using the mean ($P$ and $\dot{P}$).  The values of $\tau_c$ and $B_s$ are tabulated in Table~\ref{Table1}. These values are mostly in agreement with those in ~\citet{dirichlet}.

\setlength{\arrayrulewidth}{0.2mm}
\setlength{\tabcolsep}{6pt}
\renewcommand{\arraystretch}{0.2}
\begin{table*}[t]
\caption{\label{Table1} The Gaussian cluster parameters obtained after applying XDGMM to the natural log of pulsar $P$ and $\dot{P}$ values. These include the weights, means and covariance matrices of Gaussian clusters A, B, C, D, E and F as indicated in Fig. \ref{fig4}. Characteristic age and characteristic dipole magnetic field strength correspond to the coordinates of the cluster centre and are obtained from Eq.~\ref{eq8} and ~\ref{eq9}.}
\begin{center}
\begin{tabular}{c c c c c c}
\hline \hline
\textbf{Cluster} & \textbf{Weight} & \textbf{Mean ($P$ (sec), $\mathbf{\dot{P})}$} & \textbf{Covariance Matrix} & \textbf{Characteristic Age } & \textbf{Characteristic $B$ } \\
& & & & (Yr) & (G) \\
\\[0.5ex]
\hline 
\addlinespace[1.5ex] 
A & 0.0808 & (0.003 $0.94\times10^{-20}$) & $\begin{bmatrix}  0.0281 & 0.0063 \\ 0.0063 & 0.3064 \end{bmatrix}$ & $6.01\times10^{9}$ & $1.86\times10^{8}$  \\[3.5ex]
B & 0.0411 & (0.02 $1.94\times10^{-19}$) & $\begin{bmatrix}  0.1809 & 0.1138 \\ 0.1138 & 0.5701 \end{bmatrix}$ & $1.98\times10^{9}$ & $2.20\times10^{9}$  \\[3.5ex]
C & 0.1923 & (0.43 $2.17\times10^{-14}$) & $\begin{bmatrix}  0.2629 & -0.0384 \\ -0.0384 & 0.5035 \end{bmatrix}$ & $3.18\times10^{5}$ & $3.11\times10^{12}$  \\[3.5ex]  
D & 0.3939 & (0.61 $2.39\times10^{-15}$) & $\begin{bmatrix}  0.0999 & 0.0234 \\ 0.0234 & 0.2889 \end{bmatrix}$ & $4.02\times10^{6}$ & $1.22\times10^{12}$  \\[3.5ex]
E & 0.2750 & (0.69 $2.21\times10^{-16}$) & $\begin{bmatrix}  0.1051 & 0.1698 \\ 0.1698 & 0.7195 \end{bmatrix}$ & $4.98\times10^{7}$ & $3.96\times10^{11}$  \\[3.5ex]

F & 0.0165 & (5.51 $6.67\times10^{-13}$) & $\begin{bmatrix}  0.0443 & 0.0133 \\ 0.0133  & 1.9913 \end{bmatrix}$ & $1.31\times10^{5}$ & $6.14\times10^{13}$ \\[3.5ex]
 \addlinespace[0.5ex] 
 \hline \hline
\end{tabular}

\end{center}
\end{table*}

\subsection{Robustness of XDGMM}
~\citet{Popov} have pointed out  that the data classification using GMM does not produce stable results. 
They showed that the  exclusion of magnetars and thermally emitting neutron stars results in different clusters than the ones obtained in ~\citet{kjlee}. They also found that upon randomly removing 10\% of the data points, the positions of the resulting clusters changes significantly. 
Hence, they concluded that GMM is oversensitive to small changes in the data, and consequently cannot be used for identifying evolutionary related groups in the pulsar distribution.  

\begin{figure}[h]
\centering
\includegraphics[scale=0.38]{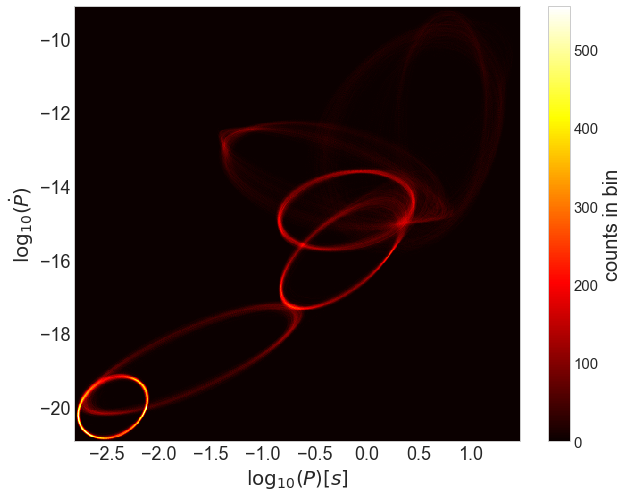}
\caption{2-D heatmap of  clusters obtained in each of our 1000 numerical experiments obtained by randomly removing 10\% of the data points.}
\label{2d_hist}
\end{figure}

To check if our XDGMM-based classification runs into similar problems, we followed the same procedure as in ~\citet{Popov}. We carried out numerical experiments, by randomly removing 10\% of the total data points in each trial run, and then applied XDGMM for different number of Gaussian clusters. After repeating this procedure
1000 times with six clusters for classification, we have used $K$-means to find the number of executions which 
resulted in a figure similar to Fig.~\ref{fig4}. We have found that, 69.2\% executions resulted in the correct output. We have plotted the 2D histogram of these output clusters as seen in Fig.~\ref{2d_hist}. 

From Fig.~\ref{2d_hist}, we find that  all the executions result in  the same clusters for MSPs, and also the cluster C, D, and E (as seen in Fig.~\ref{fig4}). Whereas, there is a clear variation in the cluster corresponding to  cluster F in Fig.~\ref{fig4} over multiple executions. Quantitatively, the centroid of this cluster varies by a maximum of 0.5 seconds (in log scale) in x-direction and 0.3 in the y-direction across multiple executions. From this, we can conclude that XDGMM cannot accurately cluster the data when the sample size is low.
 %SD: one word missing after sample size is 
\iffalse
\st{The BIC scores for different number of Gaussian clusters for the reduced data set show a similar trend as in Fig.~\ref{fig3}. The minimum value  of BIC for 6 Gaussian clusters is obtained for 90\% of our numerical experiments. Therefore, we conclude that our implementation of XDGMM is not sensitive to small changes in the dataset.}
\fi
\begin{figure}[H]
\centering
\includegraphics[scale=0.45]{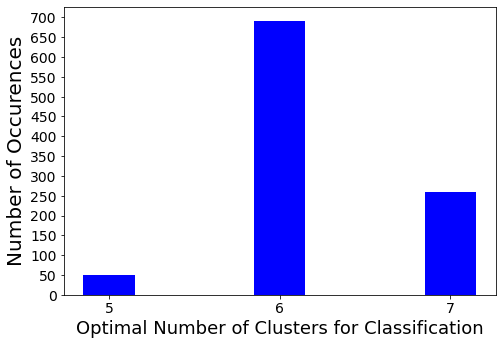}
\caption{Bar graph of the number of executions resulting in same number of optimal clusters in XDGMM.}
\label{bar_graph}
\end{figure}

We have also found the BIC scores for different number of clusters (ranging from 1-10) for all the 1000 executions. Fig.~\ref{bar_graph} shows the number of executions which resulted in 5, 6, and 7 as optimal number of clusters. We found that 69.1\% executions resulted in six, 26\% executions resulted in seven, and 4.9\% executions resulted in five as optimal number of clusters. The BIC score for executions resulting in seven clusters was very close (within 100) to the score obtained for six cluster for that particular execution. The case of 4.9\% executions resulting in five clusters as optimum could be because of the removal of 10\% data was done mainly from a single location. 

\subsection{GMM for Pulsar Classification} 
\begin{figure}
\centering
\includegraphics[scale=0.45]{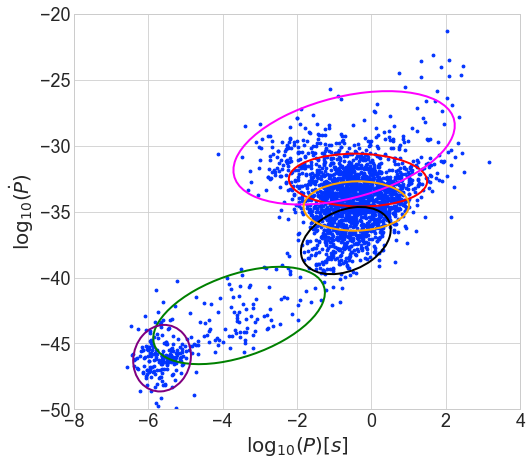}
\caption{Result 1 after applying GMM: Two clusters are needed to classify MSPs.}
\label{fig4redo}
\end{figure}

\begin{figure}
\centering
\includegraphics[scale=0.44]{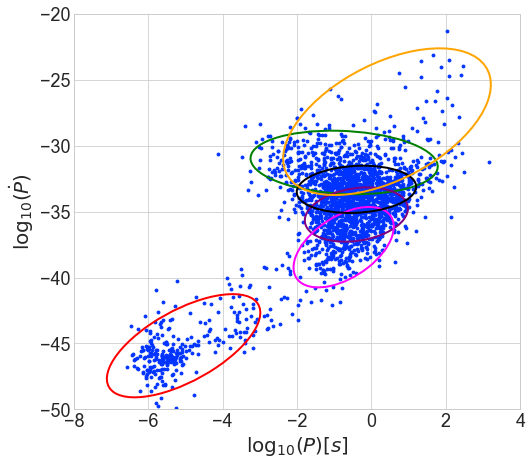}
\caption{Result 2 after applying GMM: One cluster is obtained to classify MSPs.}
\label{fig5}
\end{figure}

Given that the uncertainties for most pulsars in our dataset are negligibly small, we would like to compare our results with  GMM as a cross-check.
Therefore, we also  applied GMM to the same data described  in Sect.\ref{data} and selected the optimum number of clusters by minimizing BIC in the same way as to be done for XDGMM. We found that by applying BIC (Sec.~\ref{BIC}) to our data, the optimum number of clusters required to classify this data is six. Hence, assuming that the dataset contains six clusters, we applied GMM to this data set ten times  and found that the positions of the resulting clusters is not stable. We get a  bimodal distribution for the positions of the clusters, instead of one stable solution. The first case relates to Fig.~\ref{fig4redo}, whereas two clusters are needed to classify the MSPs and the second case relates to Fig.~\ref{fig5}, where only one cluster was needed to classify the MSPs. The probability of obtaining one of these two cases was 50\%. Therefore, we find that even without randomly removing any pulsars, we do not get a stable solution. Other problems related to application of GMM are pointed out in ~\citet{Popov}.

\begin{figure}
\centering
\includegraphics[scale=0.44]{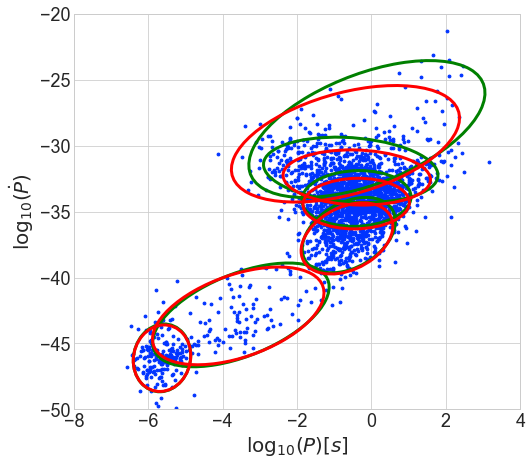}
\caption{The obtained clusters for different executions of GMM. The red clusters correspond to a single execution and the green clusters correspond to another execution.}
\label{fig6}
\end{figure}

Furthermore, even in executions which resulted in the same number of cases, it was sometimes observed that the clusters obtained were not having the same location and axes lengths. As seen in Fig.~\ref{fig6}, the position of the clusters corresponding to high energy pulsars varies for different executions.

The above observations imply that applying GMM on the Pulsar $P$-$\dot{P}$ results in a dichotomy  in  pulsar classifications with equal probability for the two groups of clusters and even in cases where two clusters were enough to classify MSPs, the position of the clusters corresponding to C, D, and F clusters (from Fig.~\ref{fig4}) is varied.
Hence, GMM cannot  be used for the classification of   pulsars in a robust fashion. 
\newline
\newline

\section{Conclusions}
\label{sec:conclusions}
Two papers within the past decade have classified the radio pulsar population along with other ancillary neutron star datasets, using unsupervised clustering techniques, such as GMM and Dirichlet mixture model~\citep{kjlee,dirichlet}. We carry out a similar exercise using XDGMM, which is an extension of GMM where the uncertainties in the observed variables are incorporated. Similar to the previous works, we carried out this classification using the logarithm  of the period and period derivative. The optimum number of clusters was chosen using the BIC criterion from information theory.

When we apply this method to the latest catalog of radio pulsars, which we obtained from the online ATNF catalog, we find that the optimum number of clusters, which can describe the radio pulsar population is equal to six. Two of these describe the millisecond pulsar population, whereas the remaining radio pulsar population can be grouped into four clusters.  However, even within each cluster, there is considerable variation in the types of neutron stars which be found and no cluster can be unambiguously associated with any specific type of pulsar or neutron star.
The 95\% confidence level ellipses showing the full dataset centered on these clusters can be found in Fig.~\ref{fig4}. The mean values, covariance matrices, the characteristic ages and magnetic fields of each of these clusters  can be found in Table~\ref{Table1}. These results are in accord with the previous works. We also tested the robustness of the XDGMM algorithm using numerical experiments and also point out some advantages compared to the ordinary GMM.

However, we should caution that this statistical classification based on only $P$ and $\dot{P}$
does not offer additional insights into pulsar phenomenology or evolution of the pulsar population, because these likely depends on additional parameters.

To promote transparency in data analysis, we have made available our codes and the data used for this analysis, which can be found at \url{https://github.com/taruntejreddych/XDGMM-for-Pulsar-classification}

\section{Acknowledgements}
We are grateful to the anonymous referee for useful constructive feedback on the manuscript.

\bibliography{main.bib}
\end{document}